\documentclass[modern,astrosymb,tighten,trackchanges]{aastex631}
\usepackage{xspace}
\usepackage{multirow}

\accepted{for publication in ApJ}
\shorttitle{Relative Sizes of Soft Excess Regions}
\shortauthors{Zoghbi et al.}

\graphicspath{{./}{figures/}}

\newcommand{\nicer}{\textit{NICER}\xspace}
\newcommand{\tons}{{TON S180}\xspace}
\newcommand{\pg}{{PG 1404+226}\xspace}
\newcommand{\oneH}{{1H0707-495}\xspace}
\newcommand{\mrk}{{MRK 335}\xspace}

\begin{document}

\title{Measuring The Soft Excess Region Size Relative to the Corona in AGN With NICER}

\email{azoghbi@umd.edu}
\author[0000-0002-0572-9613]{A. Zoghbi}
\affiliation{Department of Astronomy, University of Maryland, College Park, MD 20742}
\affiliation{HEASARC, Code 6601, NASA/GSFC, Greenbelt, MD 20771}
\affiliation{CRESST II, NASA Goddard Space Flight Center, Greenbelt, MD 20771}

\author[0000-0003-2869-7682]{J. M. Miller}
\affiliation{Department of Astronomy, University of Michigan, Ann Arbor, MI 48109, USA}

\begin{abstract}

The soft excess is a significant emission component in the Soft ($< 1$ keV) X-ray spectra of many AGN.
It has been explained by disk reflection, a warm corona and other models.
Understanding its origin is crucial for the energy budget of AGN emission, and for using it to study the inner accretion disk.
Here, we track the weeks-to-months variability of several AGN that show different levels of soft excess strength with \nicer.
We use the variability time scales to compare the relative size of the soft excess emission region to the corona producing the hard X-ray emission above 1 keV.
We find that the size of the soft excess emission region relative to the corona is not the same for the three sources studied.
For TON S180, the soft excess region is comparable in size to the hard corona.
While for MRK 335 and 1H0707-495, the soft excess region is larger than the corona by a factor of 2-4.
This is the first time the relative sizes are quantified independently of the assumptions of the spectral models.

\end{abstract}


\section{Introduction} \label{sec:intro}
A puzzling feature the X-ray spectra of AGN is the so-called soft excess. This is a strong excess of emission that is observed above the extrapolation of the power-law from the hot corona to energies below 1 keV \citep[e.g.][]{1993A&A...274..105W,2009A&A...495..421B}. The origin of this feature is still debated \citep{2019ApJ...871...88G,2020A&A...634A..85P}. It is featureless, with a shape that can generally be described by a blackbody with a temperature of $\sim$0.5 keV \citep{2004MNRAS.349L...7G}.

Early models attributed the soft excess to the tail of the disk emission \cite{1999ApJS..125..317L} or to a smeared blend absorption lines \citep{2004MNRAS.349L...7G}. In recent years, two models are often discussed in the literature. In the first, the excess is produced by the sum of recombination lines and bremsstrahlung from the heated surface of the disk that is illuminated by the hot corona. When produced close enough to the black hole, relativistic effects smear and broaden the emission lines so it appears featureless \citep{2006MNRAS.365.1067C,2013MNRAS.428.2901W,2019ApJ...871...88G}. Reverberation time lags observed in many sources are a natural consequence of this model \citep{2010MNRAS.401.2419Z,2013MNRAS.431.2441D}.

In the second model, the soft excess is produced by Comptonization of thermal disk photons by a warm ($\sim 1$ keV) and optically thick ($\tau\sim10-20$) layer of gas at the surface of the disk \citep{1998MNRAS.301..179M,2003A&A...412..317C,2012MNRAS.420.1848D,2013A&A...549A..73P}. The warm corona is distinct from the hot corona producing the primary hard X-ray emission, but they must be heated by accretion power. Both these models describe the observed time average spectra equally well, with the main discussion focusing on whether they are physically consistent \citep{2019ApJ...871...88G,2020A&A...634A..85P,2020MNRAS.496.4255B}. 

These models predict different sizes for the soft excess region. In this work, we report on a monitoring experiment with \nicer that tracks the variability of the soft excess in several sources, and use the variability to infer the size of the soft excess region relative to the corona that emits at hard energies ($>2$ keV). The combination of the large effective area and monitoring capability of \nicer allow for this experiment to be conducted for the first time.

\section{Analysis} \label{sec:analysis}
\subsection{Data} \label{sec:data}
The long term variability of 4 Narrow Line Seyfert 1 galaxies is presented. Two of which were part of the original \nicer proposal (\tons and \pg). Two others (\oneH and \mrk) that had public data on the \nicer archive were also included.

Our analysis includes all public \nicer observations available on Dec 2022. We use analysis tools available in {\sc heasoft v6.31.1}. Cleaned events files were generated from the unfiltered data using \texttt{nicerl2}. The spectra and the corresponding response and area files were then generated using \texttt{nicerl3-spect}. Background spectra were generated for every observation using all three background estimators. We find that the Scorpeon and the 3C50 models were consistent, while the space weather model over-estimates the background. We report the results from using the Scorpeon model. Using the 3C50 model gives similar resul.

\begin{figure}[ht!]
\includegraphics[width=\columnwidth,clip]{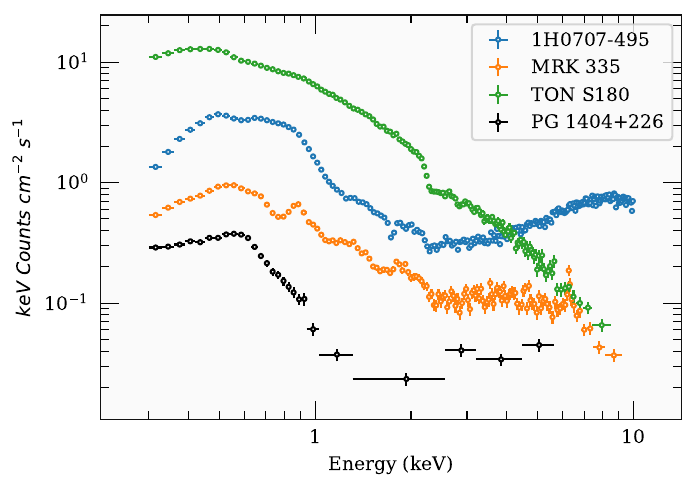}
\caption{Combined NICER spectra from all observations of each sources. These are $E F(E)$ spectra that have been unfolded against a constant model to factor out the effective area of the detector. Note that this unfolding leaves some instrumental features around 2 keV. 
\label{fig:allspec}}
\end{figure}

\begin{figure}[ht!]
\includegraphics[width=\columnwidth,clip]{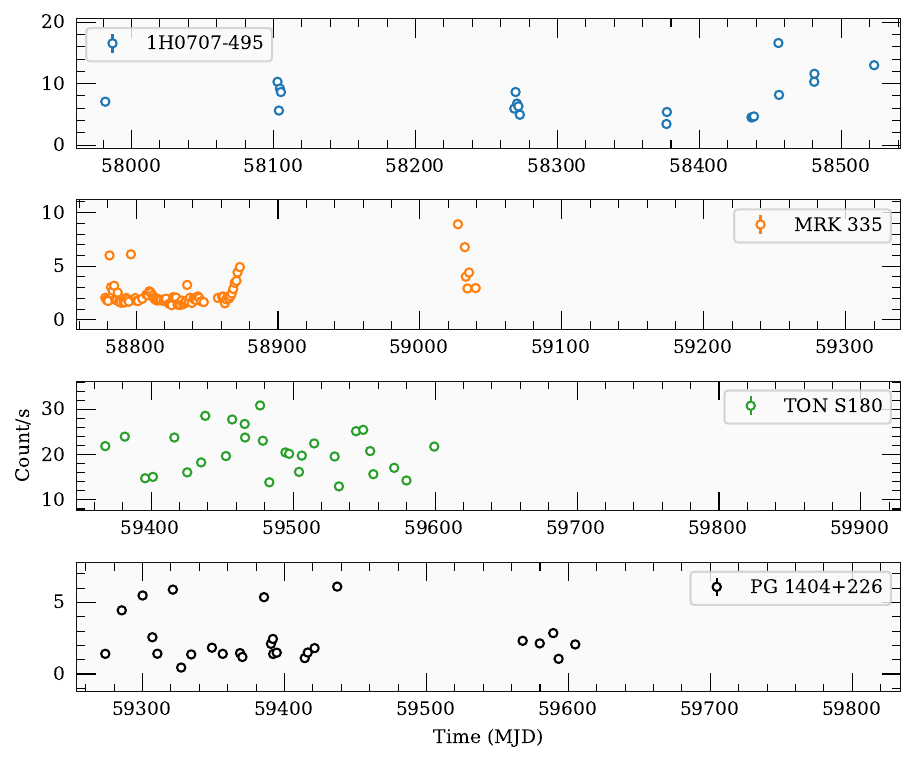}
\caption{The count rate light curves in the 0.3--10 keV band from all observations. The x-axes is in units of modified Julian date, and they all span the same $\sim580$ days.
\label{fig:alllc}}
\end{figure}

The net exposures of the resulting spectra are typically in the range of 0.6--3 ks. \oneH had a handful of observations with exposures $>10$ ks. The number of observations for each source are: 20 (\oneH), 80 (\mrk), 29 (\tons), and 26 (\pg). To visualize the spectral shape, Figure \ref{fig:allspec} shows the combined spectra for each of the 4 objects, produced by combining the individual spectra using the \texttt{addspec} tool. These are $E F(E)$ spectra that have been unfolded against a constant model to factor out the effective area of the detector. The fact that the instrument response is averaged over many months when combining the spectra results in some spurious instrument features around 2 keV. These do not affect our analysis because we are interested in the flux from broad spectral components. Figure \ref{fig:alllc} shows the total (0.3--10 keV) count rate light curves.

Figure \ref{fig:allspec} shows that the 4 objects span a range in both flux and strength of the soft excess component. Figure \ref{fig:alllc} shows that our targeted program for \tons and \pg span about a year and with some uniform sampling compared to \oneH. \mrk has the largest number of observations.  

In order to characterize the long term variability of the soft excess relative to the hard component from the corona dominating above 2 keV, we model the spectra from the individual observations with a model consisting of a power-law and a disk blackbody component. The power-law models the spectrum of the corona, dominating above 2 keV, while the disk blackbody models the soft excess. This model is used because first, we are only interested in measuring the fluxes of the two components, and second, the 0.8--3 keV does not allow for more complex models. We find that this modeling is sufficient for the purpose of measuring the fluxes needed in this work. Using other models for the soft excess have little effect on the final results.

We note that although \pg shows a strong soft excess relative the hard component (Figure \ref{fig:allspec}), the latter is too faint to enable spectral modeling of the individual spectra above 2 keV. We therefore drop it from subsequent analysis that compares the variability of the soft excess component to the hard corona.

For some observations of \oneH and \mrk, the signal above 2 keV was only high enough to constrain the flux of the power-law component, but not the spectral index. So for these two sources, we fix the spectral index of the power-law to the value measured from fitting the total spectrum with a similar model. For \tons, the signal is high enough to allow for both the flux and spectral index to be measured in the individual spectra.

After the spectral modeling of each observation, we obtain a light curve for the 0.3--10 keV fluxes for the blackbody ($F_{bb}$) and power-law $F_{p}$ components, the blackbody temperature, and for the case of \tons, for the photon index too. We then proceed by calculating and modeling the power spectral density (PSD) from these light curves.

\subsection{Power Spectra} \label{sec:psd}
The light curves are not evenly-sampled. We use the likelihood estimation method from \cite{2013ApJ...777...24Z}. We specifically use the \texttt{fqlag} package v0.3.4 \citep{abdu.zoghbi.2023.8309628}.
We first estimate the power values at 10 logarithmically-spaced frequency bins in the observed frequency range. The lowest and highest frequency limit are $1/T$ and $0.5/\Delta T_{\rm min}$, where $T$ is the length of the observing campaign, and $\Delta T_{\rm min}$ is the minimum time separation between neighboring observations. Also, as recommended by package documentation, we include two buffer frequency bins at both end of these limits to minimize biases, which are ignored when reporting the PSD.

\begin{figure}[ht!]
\includegraphics[width=\columnwidth,clip]{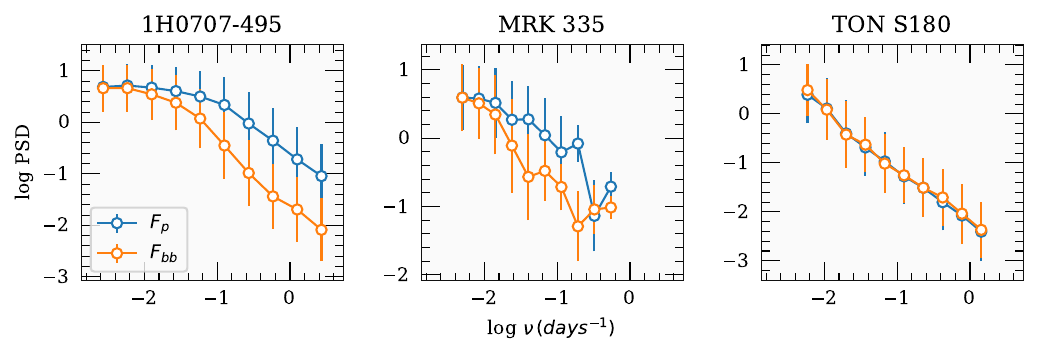}
\includegraphics[width=\columnwidth,clip]{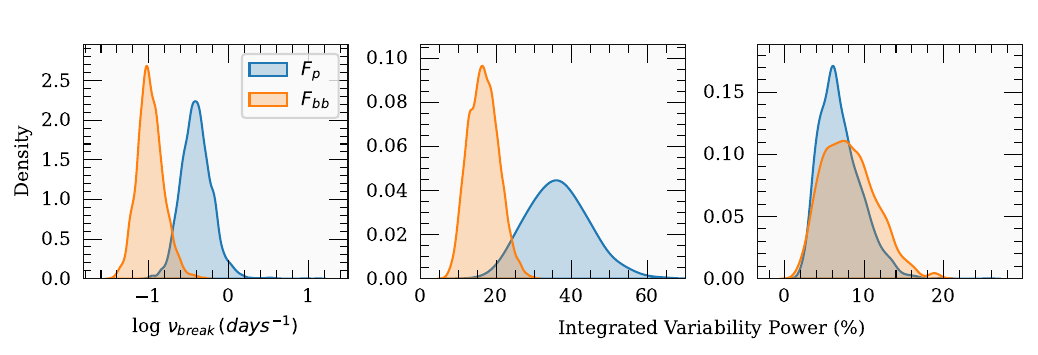}
\caption{Results of the PSD modeling for the fluxes of the blackbody ($F_{bb}$) and the coronal power-law components ($F_p$) for \oneH, \mrk and \tons. {\emph Top:} The measured PSD values at 10 frequency bins as calculated using \texttt{fqlag}. The values and errors are the median and standard deviation from the MCMC chains. The y-axis is in units of RMS-normalized PSD and x-axis is the logarithm of the the Fourier frequency. {\emph Bottom:} Smoothed density estimates for the probability density of the key parameter of interest, resulting from modeling the PSDs in the top panel by either a a zero-centered Lorentzian model (\oneH) or power-law model (\mrk and \tons). For the former case, the logarithm of the break frequency is plotted, while for the latter, the integrated variability power (integral under the power-law PSD) in percentage units is plotted.
\label{fig:psd}}
\end{figure}

Because the frequency bins may not be large enough to ensure the errors in the measured PSD values are always Gaussian, and in order to allow for subsequent fitting of the PSD, we do not just use a single value estimate of the PSD (e.g. median and a standard deviation), we instead run Monte Carlo Markov Chains (MCMC) and empirically measure the probability density of the PSD values at every frequency bin. These probability density estimates are  then approximated by a flexible standard probability density function (PDF). After trying several general functions, we found that the Johnson SB distribution provides an excellent approximation to the MCMC distributions. All the values were individually inspected to ensure the approximation is adequate. 

To characterize the measured PSDs, we fit them with three models: a power-law, a bending power-law and zero centered Lorentzian. The bending power-law has an index that smoothly changes between two values at some break frequency \citep{2004MNRAS.348..783M}. We tested fixing the lower index at both 0 and 1. Neither of them provided a significant improvement over the other two models (power-law and zero-centered Lorentzian).

In the end, we found that the power-law model was sufficient to describe the PSD of both $F_p$ and $F_{bb}$ for both \mrk and \tons, while the zero-centered Lorentzian provided a significantly better fit for \oneH. In other words, a frequency break is significantly detected only on \oneH. A plot of the PSD shapes in the three cases are shown in the top panel of Figure \ref{fig:psd}.

We note that if we use the relation between black hole mass and variability time scale from \citealt{2006Natur.444..730M}, and mass and Luminosity estimates from the literature for \mrk and \tons, we estimate the break frequencies to be: $log\nu\sim$-0.8 and 1.4 $\rm{days}^{-1}$, respectively, which are higher than our cadence allows. For \mrk, there is a hint of a break at the top of  Figure \ref{fig:psd}, consistent with the expected value, but it is not significant.

\begin{table}[]
    \centering
    \begin{tabular}{|c|c|c|c|c|}
    \hline
    \multicolumn{2}{|c|}{} & {\bf \oneH} & {\bf \mrk} & {\bf \tons} \\ 
    \hline
    
    \multirow{3}{*}{$F_p$} & $A$      &    $2.57 \pm 0.17$    &    $-0.87 \pm 0.16$     &    $-2.26 \pm 0.30$\\
    \cline{2-5} 
    
                           & $\nu_{b}$ &    $-0.41 \pm 0.19$   &    -                   &    -\\
    \cline{2-5}
    
                           & $\alpha$ &    -                   &    $-0.76 \pm 0.11$     &    $-1.22 \pm 0.25$\\
    \hline
    
    \multirow{3}{*}{$F_{bb}$} & $A$   &    $1.96 \pm 0.13$    &    $-1.37 \pm 0.15$     &    $-2.19 \pm 0.29$\\
    \cline{2-5} 
    
                           & $\nu_{b}$ &    $-0.99 \pm 0.16$   &    -                   &    -\\
    \cline{2-5}
                           & $\alpha$ &    -                   &    $-0.90 \pm 0.15$     &    $-1.23 \pm 0.20$\\
    \hline

    \end{tabular}
    \caption{Summary of the PSD fit parameters. For \oneH, a zero-centered Lorentzian is fitted to the PSD, so the amplitude $A$ (in units of RMS $\times$ days) and break frequency $\nu_{b}$ (in units of log $\rm{days}^{-1}$) are reported. For \mrk and \tons, a power-law model is fitted, so the amplitude $A$ and index $\alpha$ are reported.}
    \label{tab:fit}
\end{table}

In the bottom of Figure \ref{fig:psd}, we plot the probability distribution of a summary parameter that characterizes the variability of the flux of the two components. In the case of the \oneH, we have a direct measure of the break frequency, so we use it. For \mrk and \tons, no break in the PSD is measured, so we use the total integrated RMS variability \citep[i.e. the integral under the PSD][]{2003MNRAS.345.1271V}. For each source, we compare the variability of the blackbody ($F_{bb}$) and power-law ($F_{p}$) spectral components.

\begin{figure}[ht!]
\includegraphics[width=\columnwidth,clip]{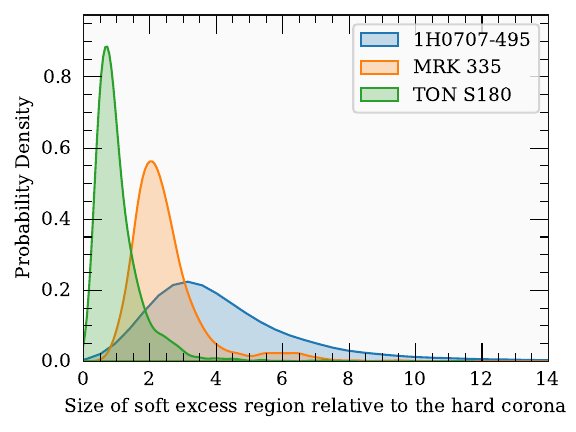}
\caption{Probability density of the relative size of the soft excess to the hot corona. The relative size is obtained as the ratio of the summary parameter for soft excess flux ($F_{bb}$) and the corona ($F_p$). For \oneH, the summary parameters is the PSD break frequency, while for the \tons and \mrk, we use the total integrated RMS variability. The $1\sigma$ single value estimates from these probability densities are: $3.7_{-1.5}^{+2.7}$, $2.2_{-0.6}^{+1.0}$ and $0.9_{-0.4}^{+1.0}$ for \oneH, \mrk and \tons, respectively.
\label{fig:relative_size}}
\end{figure}

\section{Results \& Discussion} \label{sec:results}

Section \ref{sec:psd} (summarized in Figure \ref{fig:psd}), presented one way to characterize the variability from the fluxes of the soft excess component, modeled with a blackbody, and the hard coronal component, modeled with a power-law. It is not straight forward to convert these measurement to physical units. However, we can focus on the \emph{relative} scale of the parameters for the two spectral components. 

Figure \ref{fig:relative_size} shows the ratio between the summary parameters for the soft excess component ($F_{bb}$) and the corona component ($F_{p}$). The summary parameter is the characteristic time scale (i.e. the break frequency) for the case \oneH, and the total integrated variability power in the case of the \mrk and \tons. The $1\sigma$ single value estimates from these probability densities are: $3.7_{-1.5}^{+2.7}$, $2.2_{-0.6}^{+1.0}$ and $0.9_{-0.4}^{+1.0}$ for \oneH, \mrk and \tons, respectively.

For the case of \oneH, both the break frequency and the total variability power can be measured. To check for consistency, we also measure the relative size using the variability power, we find: $4.1^{+2.3}_{-1.5}$, which is similar to the value measured from the break frequency, supporting the robustness of the result.

With the assumption that the parameter characterizing the variability (either the break frequency or the integrated RMS power) scale with the size of the emission region, the ratio in Figure \ref{fig:relative_size} maps directly to the relative size of the emission regions producing the soft excess and the hard corona. This assumption is reasonable, and is justified by the many scaling relations observed in accreting black holes. This includes the scaling of variability time-scale with black hole mass \citep{2006Natur.444..730M}, the normalized excess variance ($\sigma_{\rm NXV}^2$) scaling with black hole mass \citep{2004MNRAS.348..207P, 2012A&A...542A..83P,2022A&A...666A.127A}, and the RMS variability on long time scales in optical light curves also scaling with mass \citep{2010ApJ...721.1014M}.

Note that what we refer to as size here is the radial location of the emission region, which is equivalent to region scale size given the usual assumption of symmetry in accretion disks.

The plot in Figure \ref{fig:relative_size} shows that the soft excess region is comparable in size to the hard corona in \tons, while it is larger by a factor of 2--4 for the case of \oneH and \mrk. This estimate does not make any assumptions about the nature of the soft excess emission, and so they provide new insight into the nature of the soft excess.

\begin{figure}[ht!]
\centering
    \includegraphics[width=\columnwidth,clip]{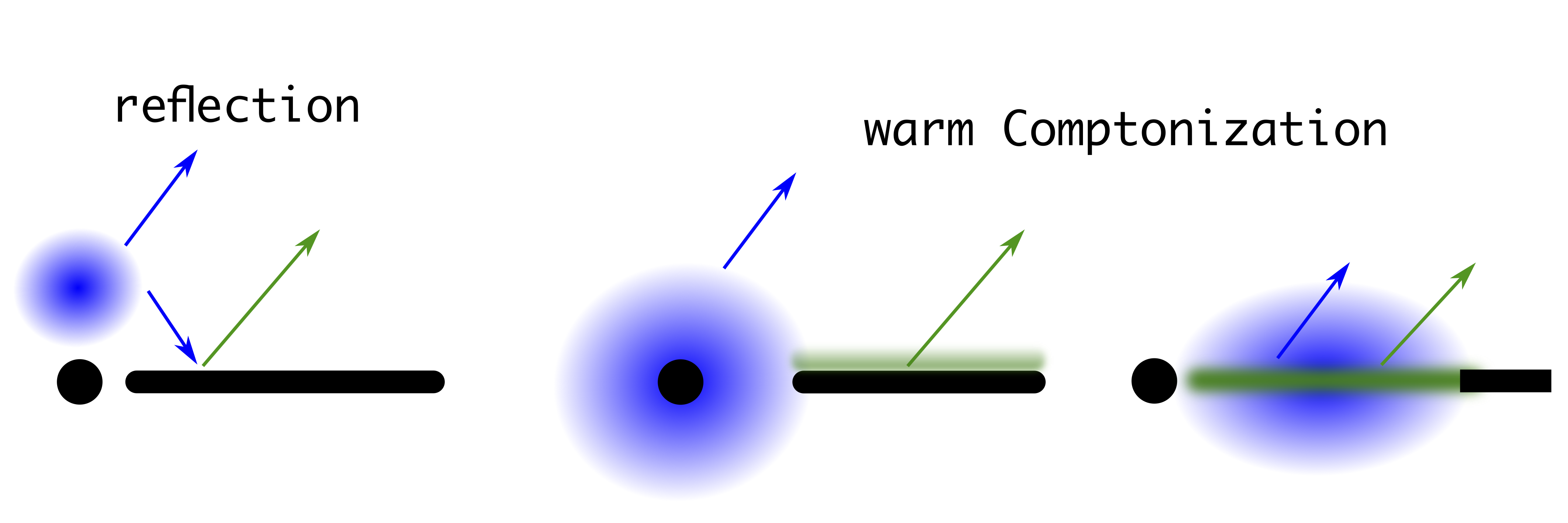} 
    \caption{\small Illustration of the different configurations that correspond to the different models proposed for the soft excess. The left scenario is for a relativistic reflection model, where the hot corona (blue) illuminates the disk producing emission lines, which when produced very close to the black hole are broadened and smeared out. The other two images are for the case of warm corona: A layer above the disk that is heated by internal dissipation that is distinct from the hot corona \citep{2013A&A...549A..72P}, and a co-located hot and warm corona that radiate inward of some radius $R_{\rm corona}$, while a standard disk emits outside it \citep{2012MNRAS.420.1848D}. }
    \label{fig:draw}
\end{figure}

A key requirement for the relativistic reflection model to explain the smooth soft excess, is that the emission has to originate very close to the black hole (only a few gravitation radii; $r_g$ from the horizon). This is needed to blur out emission lines from O and Ne that are produced below 1 keV in in a partially-ionized gas. This in turn requires the illuminating corona to be very compact \citep{2009Natur.459..540F,2010MNRAS.401.2419Z,2012MNRAS.424.1284W,2013MNRAS.428.2901W,2019MNRAS.489.3436J}. So the primary X-ray source has to be very compact (only a few $r_g$), and the reflection is also very compact (Illustrated in the left panel of Figure \ref{fig:draw}). So to a first order, the reflector is expected to be \emph{comparable} in size to the primary corona. We confirmed this by running ray-tracing simulations around a spinning black hole. Although, we leave the detailed modeling to future work, our initial simulations show that for a standard lamp-post, the relative size is undefined because the corona is assumed to be a point source. If the an extension is added to the corona, we find that the size of the reflection region (as measured for example by the radius that encompasses 90\% of the emission) is comparable to the size of the corona, consistent with emissivity profile studies \citep{2012MNRAS.424.1284W}.

The geometry for the warm corona model may not be as well defined, and it is typically an assumption in the model. In some models \citep[e.g.][]{2013A&A...549A..73P} (Illustrated in the middle panel of Figure \ref{fig:draw}), an optically-thin hot corona ($kT\sim100$ keV; $\tau\sim1$) is present in the inner parts of the accretion flow, producing the hard emission. The outer accretion flow is a vertically structured accretion disk, with cold and optically thick matter in the deeper layers, while the upper layers are composed of an optically-thick warm corona ($kT\sim1$ keV; $\tau\sim10-20$) that is powered by internal heating. The prediction from these geometries is that the soft excess region is \emph{larger} in size than the hot corona.

Other models \citep[e.g.][]{2012MNRAS.420.1848D} assume that the emission thermalizes to a (color temperature corrected) blackbody only at large radii (Illustrated in the right panel of Figure \ref{fig:draw}). At smaller radii the gravitational energy is split between powering optically thick Comptonized disk emission, forming the soft X-ray excess, and an optically thin corona above the disk, forming the hot corona tail at higher energies. Similar to the reflection case, the prediction here is that the soft excess region is comparable in size to the hot corona.

Other variants of the model \citep{2018MNRAS.480.1247K} assume an inner hot corona up to $R_{\rm hot}$, then warm corona up to $R_{\rm warm}$ and then the standard cold disk. In that specific study, a model for the spectral energy distribution can be fitted to the observations, and the different radii can be inferred. Those result suggest a relative size of $R_{\rm warm}/R_{\rm hot} \sim 2-4$ for the sources in that study, which is similar to what we measure for \oneH and \mrk. The geometry is slightly different from what we sketched in Figure 5, but it would fall under the middle sketch where the warm corona is outside the hot corona.

Our results suggest that for \tons, where the size of the soft excess region is comparable to the hot corona, the relativistic reflection model or the warm corona that sandwiches a hot corona are possible geometries. On the other hand, for \oneH and \mrk, a warm corona that is larger than the hot corona appears to be more consistent with the data. 

In our discussion of the relative sizes, we are ignoring the detailed geometries (e.g. spherical vs flat) and viewing angle. For a simple spherical emission regions, there is only scale size that controls  the variability. For other shapes, say a torus, there are different scales in different directions, but the variability should be dominated by the largest scale, regardless of our viewing angle. Consequently, the flux may depend on the angle, but the variability timescale does not.

We plan to continue monitoring these and other sources with a soft excess to obtain further constraints.

\noindent
The code used to produce these results is available on Zenodo \citep{zoghbi.abdu.2023.8310123}. The data products are available on the Open Science Framework site \citep{Zoghbi.data.2023}.

\begin{acknowledgments}
The material is based upon work supported by NASA under award numbers 80GSFC21M0002 and 80NSSC23K0333.
\end{acknowledgments}

%

\vspace{5mm}
\facilities{NICER(XTI), HEASARC}





\bibliographystyle{aasjournal}



\end{document}